# Spin-Orbit Torque Characterization in a Nutshell

Minh-Hai Nguyen[a]

*Raytheon BBN Technologies, Cambridge, Massachusetts 02138, USA*

Chi-Feng Pai[a]

*Department of Materials Science and Engineering, National Taiwan University, Taipei 10617, Taiwan*

Spin current and spin torque generation through the spin-orbit interactions in solids, of bulk or interfacial origin, are at the heart of spintronics research. The realization of spin-orbit torque (SOT) driven magnetic dynamics and switching in diverse magnetic heterostructures also paves the way for developing SOT magnetoresistive random access memory (SOT-MRAM) and other novel SOT memory and logic devices. Of scientific and technological importance are accurate and efficient SOT quantification techniques which have been abundantly developed in the last decade. In this article, we summarize popular techniques to experimentally quantify SOTs in magnetic heterostructures at micro- as well as nano-scale. For each technique, we give an overview of its principle, variations, strengths, shortcomings, error sources and any cautions in usage. Finally, we discuss the remaining challenges in understanding and quantifying the SOTs in heterostructures.

## I. Introduction

The concept of current-induced spin-polarization in non-magnetic solids can be dated back to the 1970s, when two theoretical physicists Dyakonov and Perel proposed that a longitudinal charge current applied in a thin film semiconductor with spin-orbit interaction will lead to the reorientation of the electron spins.[1] This transverse spin accumulation or spin current generation by applying a longitudinal charge current in paramagnetic materials was re-predicted by Hirsch in 1999, and coined as the spin Hall effect (SHE).[2] Phenomenologically the SHE can be expressed as

$$\boldsymbol{j}_s = \left(\frac{\hbar}{2e}\right)\theta_{\text{SH}}(\hat{\sigma}\times\boldsymbol{j}_e), \quad (1)$$

where $\boldsymbol{j}_e$, $\boldsymbol{j}_s$, and $\hat{\sigma}$ represent the applied longitudinal charge current density, induced transverse spin current density, and electron spin-polarization unit vector, respectively, as illustrated in Fig. 1(a). The spin Hall angle $\theta_{\text{SH}}$, whose magnitude and sign depend on the material, parametrizes the strength of the SHE. Note that by the original definition of spin-dependent Hall angle, it should be $\tan\theta_{\text{SH}}$ in eq. (1) instead of $\theta_{\text{SH}}$.[3] However, an approximation of $\tan\theta_{\text{SH}} \approx \theta_{\text{SH}}$ has been used since this angle was believed to be small. This condition might not hold if the spin-dependent deflection is significant, as we shall see below for materials with large spin-orbit interactions. (Nevertheless, we keep the conventional usage of $\theta_{\text{SH}}$ throughout this article.)

---

[a] Authors to whom correspondence should be addressed: minh-hai.nguyen@rtx.com and cfpai@ntu.edu.tw

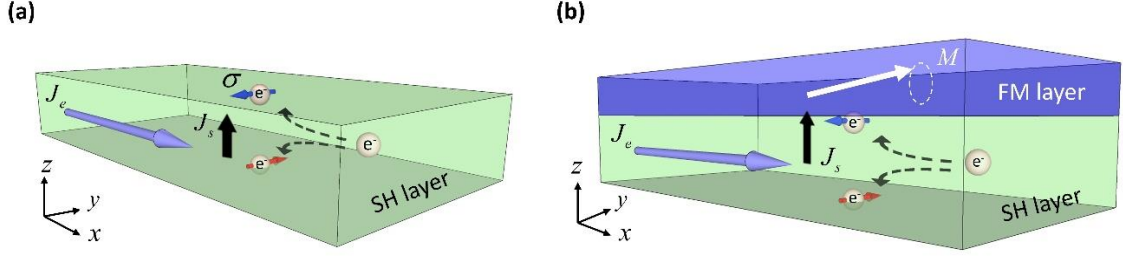

Figure 1: Illustration of the SHE in (a) a SH layer and (b) a SH/FM heterostructure.

In 2004, the SHE was first experimentally observed through optical means in semiconductors, in which the out-of-plane spin polarizations were detected at the edges of a microstrip device while the charge current was applied longitudinally.[4] Later in 2006, Valenzuela and Tinkham reported on the first electrical characterization of the SHE in Al through nanoscale nonlocal spin injection and inverse SHE (ISHE) voltage measurement.[5] $\theta_{SH}$ of these spin Hall (SH) layers are at most around the order of $10^{-4}$, therefore hard to quantify. It was not until the discovery of the sizable SHE in 5d transition metal Pt[6] that the topic became the center of spintronics research ($\theta_{SH}$ of Pt has been quantified to be $\sim 10^{-4}$ to $10^{-2}$, depending on characterization technique[7]).

More importantly, if a ferromagnetic (FM) layer is placed adjacent to the SH layer to form a SH/FM heterostructure, $j_s$ can be transmitted to the FM layer to affect the magnetization dynamics in FM, as illustrated in Fig. 1(b). In the early 2010s, the discovery of such current-induced magnetization switching in Pt-based magnetic heterostructures[8,9] and Ta-based three-terminal magnetic tunnel junction (MTJ) devices[10] eventually led to the proliferation of the so-called spin-orbit torque (SOT) research, since the spin-torques (STs)[11,12] that drive such magnetization dynamics are of spin-orbit interaction origins, either bulk (SHE) or interfacial (Rashba-Edelstein effect). Without arguing about the bulk and interfacial origin of the SOTs, in this article we use SHE and SOT interchangeably. Till this day, the family of SOT material systems has been extended from conventional transition metals to the realm of emergent materials, such as topological insulators (TIs) and transition metal dichalcogenides (TMDs). Diverse SOT characterization techniques have been proposed to quantify $\theta_{SH}$ or SOT efficiencies in this broad range of materials, heterostructures, and devices.

The reason behind the multitude of SOT quantifying techniques lies in the fact that the SOTs cannot be measured directly within the SOT materials but instead in their effects on the magnetization dynamics of the adjacent FM layer. Therefore, interfacial (*e.g.* spin transmission, spin memory loss, or spin-orbit scattering across the SH/FM interface) and FM bulk (*e.g.* magnetoresistance effects) factors need to be considered. In each technique, some factors are considered and the others assumed to be negligible, depending on their working principles. Each technique also requires a different set of material parameters whose values are well determined for some but not so for others. When not fully comprehended, these differences in measuring techniques can cause controversies of the results.[7] It is therefore of utmost importance for researchers in the SOT community to have a firm understanding of various SOT quantification techniques, their strengths and weaknesses, and how to mitigate errors.

Although several detailed reviews have been given to cover the physics and the device aspects of the SHE[13] and SOT,[14,15] an overview and in-depth discussion on the diverse types of SOT characterization techniques is still lacking. In the following sections, we will first walk through seven most popular techniques that have been employed to characterize the spin Hall angle or the SOT efficiencies in magnetic heterostructures. For each one, we will go over its working principle, variations, advantages and disadvantages, major error sources and any cautions in usage. We then summarize and compare the strengths and limitations of these techniques. A prospect of future development of SOT devices and discussion on the remaining challenges, especially those regarding characterization, are given in the final section.

## II. Models and assumptions

Before surveying the techniques to quantify the SOTs, we need to understand what exactly we are and are not measuring. From the experimental point of view, it is near impossible to quantify the SOTs directly at the source at atomic level. Instead, we normally measure their effects on the magnetization dynamics of the adjacent FM layer. We have to make a series of assumptions to reach a reasonably simple model whose parameters can be extracted by experiments. First of all, we describe the magnetic states of the sample by a single magnetization vector $\boldsymbol{M}$, which is often normalized as $\boldsymbol{m}$ by its saturation magnetization $M_s$. This *macrospin* picture also assumes that the magnetization dynamics of the sample is uniform, consequently inducing the next approximation: the effects of external fields and torques on the sample are uniform and represented by their corresponding vectors acting on $\boldsymbol{M}$, although in reality they are stronger near the interface with the SH layer and more complicated along the edges.

At the point of this writing, the community has largely agreed on the two distinct SOT effects on the magnetization in a SH/FM heterostructure: the damping-like (DL) and field-like (FL) effects, corresponding to the so-called DL and FL SOT[16]. Consider a macrospin magnetization $\boldsymbol{m}$ under an applied field $\boldsymbol{H}_{\text{ext}}$ and SOTs $\tau_{\text{DL}}$ and $\tau_{\text{FL}}$. Its dynamics can be described by the general Landau-Lifshitz-Gilbert-Slonczewski equation:

$$\frac{d\boldsymbol{m}}{dt} = -\gamma \boldsymbol{m} \times \boldsymbol{H}_{\text{ext}} + \alpha \boldsymbol{m} \times \frac{d\boldsymbol{m}}{dt} - |\tau_{\text{FL}}|\boldsymbol{m} \times \hat{\sigma} - |\tau_{\text{DL}}|\boldsymbol{m} \times (\boldsymbol{m} \times \hat{\sigma}), \qquad (2)$$

where $\gamma$ is the gyromagnetic ratio and $\alpha$ the Gilbert damping coefficient. In many techniques reviewed in this article, we don't measure the SOTs directly but model them as *effective fields*. This can be done by replacing the two terms $|\tau_{\text{DL}}|(\boldsymbol{m} \times \hat{\sigma})$ and $|\tau_{\text{FL}}|\hat{\sigma}$ in eq. (2) by the effective fields $\gamma \boldsymbol{H}_{\text{DL}}$ and $\gamma \boldsymbol{H}_{\text{FL}}$. This conversion of torques into effective fields allows us to use energy conservation law to build simplified analytical expressions for fitting experimental data. Note that this approximation is acceptable only if their effects on $\boldsymbol{m}$ are so small that $\boldsymbol{m}$'s movement is confined within a small angle ~10°. When $\boldsymbol{m}$ moves in a large angle, as in current-induced magnetization switching or oscillation, it is no longer valid due to the fact that the effect of (anti-)damping SOT on $\boldsymbol{m}$ is non-conservative while that of its effective field conservative.

Finally, what we observe experimentally is the response of $\boldsymbol{m}$ in the FM layer due to the excitation of the externally applied fields and spin accumulation at the interface with the SH layer.

Cautions have to be taken when converting the extracted effective fields to the SH layer's SOT properties: the relation between the electrical current density in the SOT layer $j_e$ and its induced effective fields $H_{DL}$ and $H_{FL}$ in the FM layer is represented by the damping-like (DL) and field-like (FL) *effective spin Hall angles* or *spin torque efficiencies* $\xi$ as:

$$\xi_{DL(FL)} = 2K_{pre}t_{FM}H_{DL(FL)}/j_e, \quad (3)$$

where $t_{FM} = t_{FM}^{nominal} - t_{FM}^{dead}$ is the (magnetic) thickness of the FM layer and the prefactor $K_{pre} = e\mu_0 M_s/\hbar$. The SH layer's internal spin Hall angle(s) can be estimated only with the knowledge of the spin transmission and spin memory loss at its interface with the FM layer, and with the assumption that the electrical and spin currents are uniform which is reasonable for SH films a few times thicker than its spin diffusion length and mean free path. Typically the attenuation factors of the spin transmission at the interface are absorbed into a complex parameter $T_{int}$, which relates the effective ST efficiencies $\xi_{DL(FL)}$ to the internal spin Hall angle $\theta_{SH}$ as $\xi_{DL} = \theta_{SH}\text{Re}(T_{int})$ and $\xi_{FL} = \theta_{SH}\text{Im}(T_{int})$. Within the spin drift-diffusion model,[16,17] $T_{int}$ is given as[18]

$$T_{int} = \frac{2G_{mix}}{2G_{mix} + (\sigma_{SH}/\lambda_{SH})\tanh(t_{SH}/\lambda_{SH})} \tanh(t_{SH}/\lambda_{SH})\tanh(t_{SH}/2\lambda_{SH})$$
$$\approx \frac{2G_{mix}^{eff}}{(\sigma_{SH}/\lambda_{SH})}\tanh(t_{SH}/2\lambda_{SH}), (4)$$

where $\lambda_{SH}$ is the spin diffusion length, $\sigma_{SH}$ the electrical conductivity of the SH layer, $G_{mix}$ the (complex) spin-mixing conductance (usually calculated theoretically), and $G_{mix}^{eff}$ the effective spin-mixing conductance whose real value can be estimated by measuring $\alpha$ as $\text{Re}(G_{mix}^{eff}) = \mu_0 M_s t_{FM}(\alpha - \alpha_0)/(\gamma\hbar)$ ($\alpha_0$: intrinsic damping coefficient, can be estimated by measuring $\alpha$ of FM without the SH layer).

Keeping in mind the above approximations, we now proceed to discuss various methods to quantify the SOT effects on the dynamics of ***m*** in the FM layer.

## III. Nonlocal spinvalve

One of the earliest techniques to quantify the DL SOT efficiency through electrical means is the nonlocal spinvalve (NLSV) approach.[5] A full description of the method can be found in Niimi & Otani[19] and its full derivation by Kimura *et al.*[20] A modified version of NLSV can be used to measure the spin diffusion length, but that is beyond the scope of this article. The basic NLSV, described below, measures the electrical-spin conversion efficiency when an electrical current is applied in a SH nanowire and the SHE-induced spin current is detected in a FM nanowire (or vice versa), connected by a good conductor.

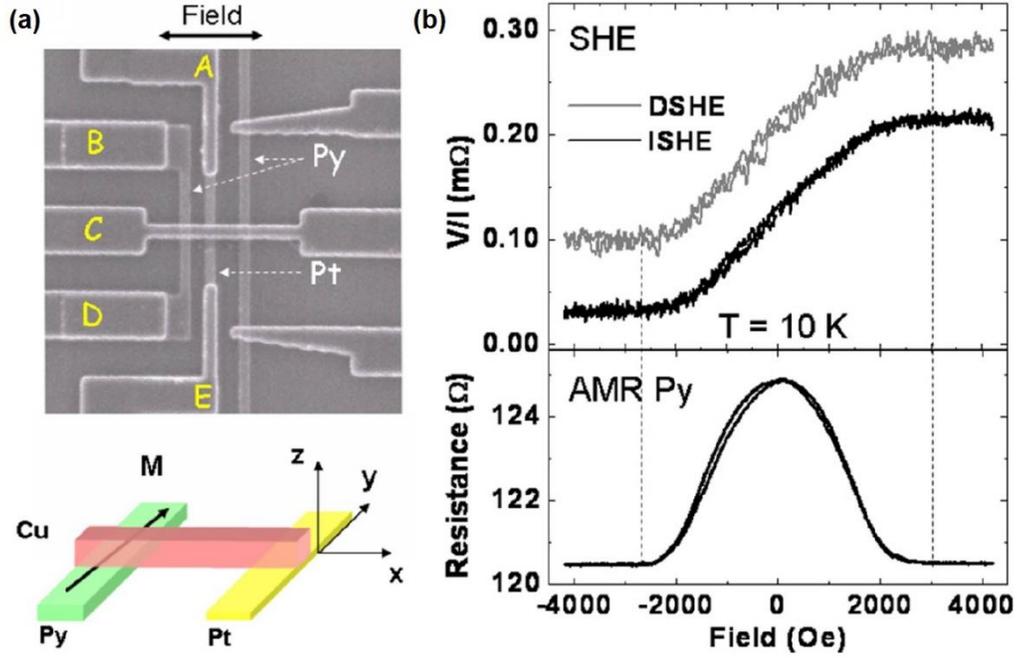

Figure 2: NLSV technique (a) SEM image and illustration of a typical device. (b) Direct and Inverse SHE (DSHE and ISHE) traces and the AMR from the Py wire as functions of the applied field. DSHE measurement corresponds to $V_{BC}/I_{AE}$, and ISHE to $V_{EA}/I_{BC}$ (A, B, C and E are the contact leads as denoted in the SEM image). Courtesy of Vila et al.[21]

As depicted in Fig. 2, the sample consists of a SH nanowire connected to a FM nanowire via a thin bridge (N) made of good electrical and spin conductor (usually Cu for its long mean free path and spin diffusion length). An in-plane field is applied along the hard axis of the FM wire (along $x$ direction in Fig. 2(a)). There are two ways to quantify the DL SOT efficiency in this configuration:

(i) *Direct SHE*: a current is applied to the SH wire, induces a spin current into the bridge via SHE. The spin current is then absorbed at the FM wire, causing a voltage change.

(ii) *Inverse SHE*: a current is applied to the FM wire, induces a spin current into the bridge.[22,23] The spin current is then absorbed at the SH wire, causing a voltage change due to the inverse SHE (the reciprocal counterpart of SHE).

As shown in Fig. 2(b), the voltages measured by the two ways are equivalent, clearly demonstrating the Onsager reciprocality. In the Inverse SHE measurement, the DL SOT efficiency is determined as[19]

$$\xi_{DL} = \frac{w_{SH}\Delta V}{(\rho_{SH}-\rho_N)I_s}, \quad (5)$$

where $\Delta V$ the voltage difference with and without the applied current; $w$, $\rho$ denote the width and resistivity; and $I_s$ is the effective spin current injected into the SH wire, which is calculated as

$$I_S = \frac{1}{\delta_{SH}} \frac{(1-e^{-\delta_{SH}})^2}{1-e^{-2\delta_{SH}}} \frac{2\rho_{FM}I_e Q_{FM}[\sinh l + Q_{FM}e^l]}{[\cosh l - 1] + 2Q_{SH}\sinh l + 2Q_{FM}[(1+Q_{FM})(1+2Q_{SH})e^l - 1]} \quad , (6)$$

where $I_e$ is the applied electrical current, $\delta = t/\lambda$ ($t$: thickness, $\lambda$: spin diffusion length), $l = L/2\lambda_N$ ($L$: the separation of the SH and FM wires), $Q_{FM(SH)} = R^s_{FM(SH)}/R^s_N$ in which $R^s = \rho\lambda/[(1-\rho^2)A]$ is the spin resistance, and $A$ the effective cross section for the pure spin current ($A_N = w_N t_N$, $A_{FM} = w_N w_{FM}$, $A_{SH} = w_N w_{FM} \tanh \delta_{SH}$).

Although this technique was developed very early, its proliferation has been limited due to a few challenges: the determination of material parameters, fabrication and noise. As seen in equation (6), the calculation of $\xi_{DL}$ requires good estimation of the film thicknesses, resistivities and especially spin diffusion lengths. The determination of $t_{SH}$ may not be as straightforward as it seems. If the underlayer is not prepared well, the electrical conductivity of the SH layer near the underneath layer can be very small, leading to an electrically dead layer. For example, from our experience, adding a thin Ta film between the SiO2 substrate and Pt film reduces the Pt film's dead layer from roughly 1 nm to below 0.5 nm.

Note that $I_e$ is the current in the SH or FM wire (for direct or inverse SHE measurement, respectively). It is related to the total applied current via a shunting factor whose accuracy depends on the estimation of resistivities, which is not quite straightforward in thin films due to surface scattering.[24] The calculation of $\xi_{DL}$ is also sensitive to the spin diffusion length of the SH material whose value for Pt has long been controversial.[7,25]

Equation (6) is derived based mainly on the one-dimensional spin drift diffusion along the bridge, treating the FM and SH wires as infinitesimally narrow so that the spreading of spin current near the interfaces of the bridge and the FM and SH wires is negligible. A more comprehensive treatment would go beyond analytical expressions and require 2D or 3D finite-element simulation.[26] To avoid error due to this approximation, the FM and SH wires need to be very narrow, typically about 100 nm. This requirement leads to two challenges: costly fabrication and higher noise. As seen from Fig. 2(b), the ratio $\Delta V/I$ is about 0.2 mΩ, hence $\Delta V$ about 200 nV for a current of 1mA, equivalent to the thermal noise of a 50 kΩ resistor at room temperature. Thus, there is an upper limit of the number of squares of the SH and FM wires within which the voltage signal can be reliably detected. The thermal noise can be remedied by measuring the sample at cryogenic temperatures, but this adds a substantial cost and introduces extra complexities (temperature-dependent parameters) to the experiment.

## IV. Spin pumping

Spin pumping technique was introduced by Saitoh *et al.* in 2006[27] to quantify the spin Hall angle via the inverse spin Hall effect (ISHE). The experiment is very similar to a typical FMR measurement, except that the sample is a FM/SH bilayer microstrip and the read out is its voltage. The sample can be placed in a microwave cavity[27] or underneath a coplanar waveguide (CPW)[28] to induce FMR. By analyzing the voltage drop as a function of the applied magnetic field (or frequency), a number of important parameters can be extracted: the sample's effective field (caused by its magnetization and anisotropy), Gilbert damping coefficient and DL ST efficiency.

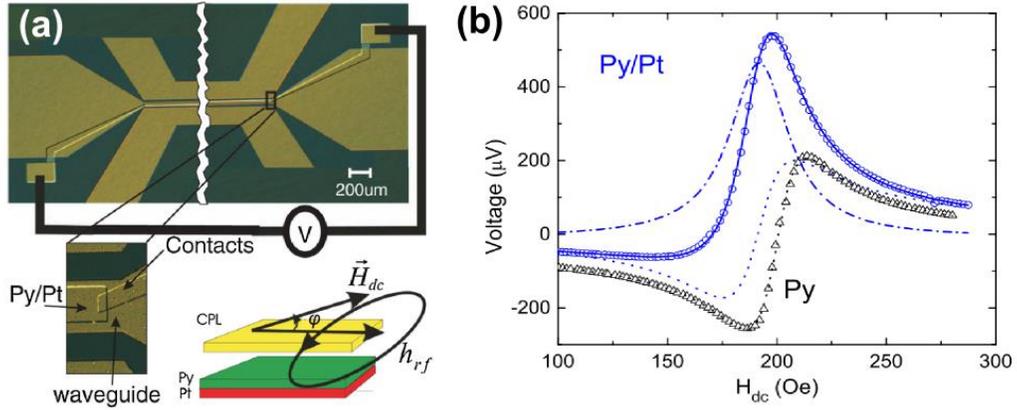

Figure 3: Spin pumping technique. (a) Experiment setup: optical image and schematic of a Py/Pt sample on a coplanar waveguide under applied DC magnetic field. (b) FMR spectra of a Pt/Pt (circles) and Py (triangles) sample. Lines are fits to the Lorentzian FMR functions. Blue dashed and dotted lines show the symmetric and asymmetric components of the Pt/Pt spectrum. Courtesy of Mosendz *el al.*[28]

A typical experiment setup is depicted in Fig. 3(a): The sample is a FM/SH bilayer microstrip placed along but electrically isolated from a CPW. A magnetic field $H_{\text{ext}}$ is applied at the angle $\phi$ with the CPW. RF excitation of frequency $f$ is applied to the CPW and the microstrip's homodyne DC voltage drop is measured. The precessing magnetization in the microwave-excited FM induces a spin current pumped into the SH layer. This spin current in turns induces a transverse electrical current in the SH layer via the ISHE. The FMR spectrum (microstrip's voltage) consists of two components: a symmetric ISHE and an asymmetric AMR voltage,

$$V_{\text{mix}}(H_{\text{ext}}) = V_{\text{ISHE}} \frac{\Delta^2}{\Delta^2+(H_{\text{ext}}-H_0)^2} + V_{\text{AMR}} \frac{\Delta(H_{\text{ext}}-H_0)}{\Delta^2+(H_{\text{ext}}-H_0)^2}, \qquad (7)$$

where $\Delta$ is the spectrum linewidth and $H_0$ the resonance field. Typical FMR spectra for Py/Pt and Py and their decomposed contributions are shown in Fig. 3(b). The DL ST efficiency can be calculated from the symmetric $V_{\text{ISHE}}$ as

$$V_{\text{ISHE}} = -\xi_{\text{DL}} C_1 C_2 \frac{efL}{2t_{\text{SH}}} \sin\phi \sin^2\Phi, (8)$$

where $L$ is the length of the sample, $\Phi$ the precession cone angle, $C_1 = \sigma_{\text{SH}} t_{\text{SH}}/(\sigma_{\text{SH}} t_{\text{SH}} + \sigma_{\text{FM}} t_{\text{FM}})$ the shunting factor ($\sigma$: electrical conductivity), and

$$C_2 = 2\omega \frac{\gamma\mu_0 M_s + \sqrt{(\gamma\mu_0 M_s)^2 + 4\omega^2}}{(\gamma\mu_0 M_s)^2 + 4\omega^2} \qquad (9)$$

is the correction factor of the precession trajectory ($\omega = 2\pi f$).[29]

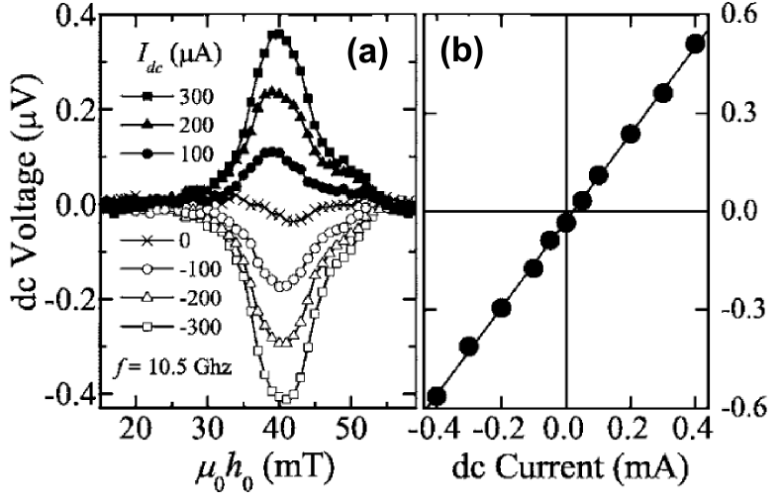

Figure 4: Determination of precession angle. (a) AMR voltage change due to the opening angle of the magnetization precession at FMR and (b) AMR maxima versus DC bias current. The solid line is linear fit to the data. Courtesy of Costache *et al.*[30]

A critical parameter in the estimation of $\xi_{DL}$ based on equation (8) is the precession cone angle $\Phi$. It can be estimated by a separate measurement proposed by Costache *et al.*[30]: set the external field along the sample ($\phi = 0$), apply a DC current $I_{DC}$ to the bilayer sample, excite FMR and measure the AMR voltage $V_\Phi$. At resonance, the precession cone angle opens up, resulting in an AMR voltage drop $V_\Phi = I_{DC} \Delta R_{AMR} \sin^2 \Phi$ where $\Delta R_{AMR}$ is the maximum change of AMR resistance (between $\phi = 0$ and $\phi = \pi/2$). Typical AMR traces and their maxima versus $I_{DC}$ for this experiment are shown in Fig. 4.

One of the advantages of the spin pumping technique is its versatility: the same experiment can be used to estimate several important parameters such as the DL ST efficiency $\xi_{DL}$, the Gilbert damping coefficient (from FMR linewidth vs. frequency) and the effective demagnetization field (from the resonance field vs. frequency followed by fitting to Kittel formula[31]). Another advantage is its ability to measure small $\xi_{DL}$ by increasing the sample length $L$ as implied by equation (8). The sample size used for this technique can be in the orders of micrometers or even millimeters, which can be readily fabricated.

The errorbar of $\xi_{DL}$ estimated by this technique comes from two main sources: the estimations of the FM and SH layers' conductivity and thickness, as discussed in the previous section (NLSV); and the measurement of the precession cone angle $\Phi$. From equation (8), for small angle $\Phi$, $V_{ISHE} \propto \xi_{DL} \sin^2 \Phi \approx \xi_{DL} \Phi^2$, thus $\Phi$'s errorbar must be reduced as much as possible. It is also important to note that the detected rectification signals can be affected by the artefacts originating from the geometry of the setup as well as the relative phase between the induced microwave current and the magnetic field,[32,33] therefore extra cares must be taken to provide a trust-worthy estimation of $\xi_{DL}$ from the measured spectrum.

## V. Spin-torque ferromagnetic resonance

There is no doubt that the ST-FMR technique had revolutionized the study of the SHE in metals. Since its first development by Liu *et al.* in 2011,[9] the technique had been quickly and widely adopted by many research groups for its simplicity and versatility. As depicted in Fig. 5(a), the experiment involves an in-plane magnetized microstrip of SH/FM bilayer under an external field $H_{\text{ext}}$ at angle $\phi$ (typically 45 degrees) to an applied RF current of amplitude $I_{\text{rf}}$ and frequency $f$. From the DC voltage response versus field strength with varying frequency, the DL ST efficiency, Gilbert damping coefficient and effective demagnetization field can be extracted.

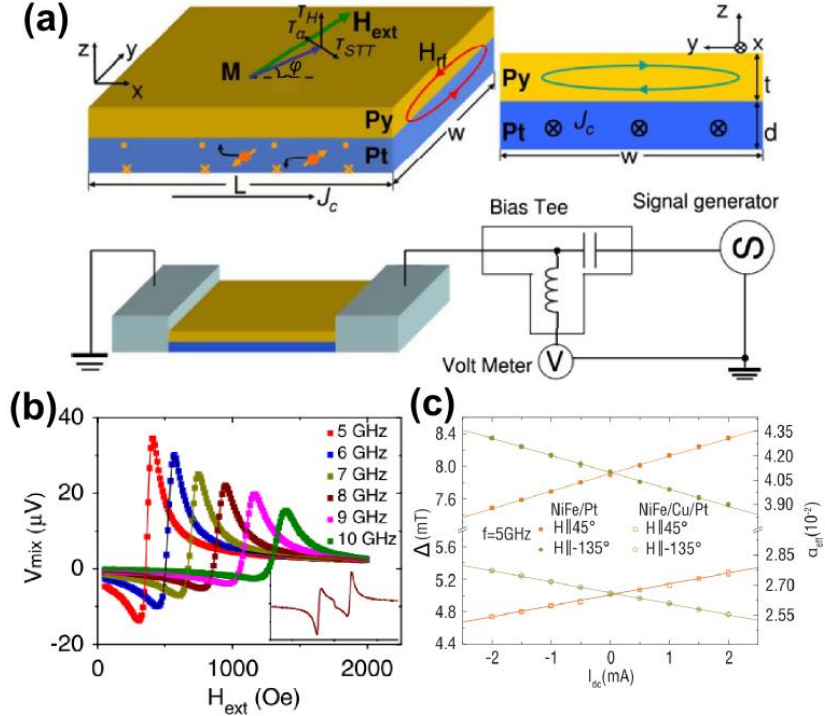

Figure 5: ST-FMR method. (a) Measurement schematic. (b) Typical ST-FMR traces at different driving frequencies of a NiFe/Pt sample. Courtesy of Liu *et al.*[9] (c) ST-FMR linewidth and (effective) damping coefficient as functions of DC bias current. Courtesy of Nan *et al.*[34]

In the original version,[9] the voltage response to RF applied current has the form (typical traces shown in Fig. 5(b))

$$V_{\text{mix}}(H_{\text{ext}}) = S\frac{\Delta^2}{\Delta^2+(H_{\text{ext}}-H_0)^2} + A\frac{\Delta(H_{\text{ext}}-H_0)}{\Delta^2+(H_{\text{ext}}-H_0)^2}, \quad (10)$$

where $S$ and $A$ are the coefficients of the symmetric and antisymmetric response to $H_{\text{ext}}$. Since $S$ is proportional to the strength of the DL spin torque, and $A$ the Oersted field induced by the RF current, the DL ST efficiency can be calculated from their ratio as

$$\xi_{\text{FMR}} = K_{\text{pre}} t_{\text{FM}} t_{\text{SH}} \frac{S}{A} \sqrt{1 + H_{\text{eff}}/H_{\text{ext}}}, \qquad (11)$$

where $H_{\text{eff}}$ is the effective demagnetization field which can be determined by fitting $H_0$ versus $f$ to Kittel formula[31].

The advantages of the lineshape ST-FMR method are readily seen: the micron-sized sample is easily fabricated with standard photolithography; it is self-calibrated in the sense that most parameters required to compute $\xi_{\text{DL}}$, except $M_s$ and film thicknesses, can be extracted from the same experiment. The Gilbert damping coefficient $\alpha$ can also be estimated by linear fitting linewidth $\Delta$ versus frequency $f$ as

$$\Delta = \frac{2\pi f}{\gamma}\left(\alpha + \xi_{\text{DL}} \frac{\sin\phi}{K_{\text{pre}} t_{\text{FM}}(2H_{\text{ext}} + H_{\text{eff}})} J_{\text{SH,dc}}\right), \quad (12)$$

where $J_{\text{SH,dc}}$ the DC bias current density in the SH layer. Equation (12) above reveals another way to estimate $\xi_{\text{DL}}$ by linear fitting $\Delta$ versus $J_{SH,dc}$ (representative data in Fig. 5(c)).[9,34] The simplicity and versatility of ST-FMR makes it a neat and popular technique in the SOT community.

It took a few years until a missing factor of the original ST-FMR method was realized: the existence of a FL effect accompanying its DL counterpart was widely accepted by the SHE community and therefore correction had to be made to ST-FMR. Because the FL torque acts on the magnetization $\boldsymbol{m}$ and hence on $V_{\text{mix}}$ the same way as the Oersted field, the observed $\xi_{\text{FMR}}$ in equation (11) is no longer identical to $\xi_{\text{DL}}$ but rather determined by both $\xi_{\text{DL}}$ and $\xi_{\text{FL}}$ as[18]

$$\frac{1}{\xi_{\text{FMR}}} = \frac{1}{\xi_{\text{DL}}}\left(1 + \frac{\xi_{\text{FL}}}{K_{\text{pre}} t_{\text{FM}} t_{\text{SH}}}\right), \quad (13)$$

which requires repeated measurements on samples having different thickness $t_{\text{FM}}$ or $t_{\text{SH}}$. To avoid variation in multiple film growths, a good way to vary $t_{\text{FM}}$ or $t_{\text{SH}}$ is wedge sputter deposition in which the sputtering gun is off from the center of the stationary wafer, while keeping the other films uniform. Note that $t_{\text{SH}}$ and $t_{\text{FM}}$ have to be thick enough so that $\xi_{\text{DL}}$ and $\xi_{\text{FL}}$ are essentially unchanged because equation (13) assumes that $\xi_{\text{DL}}$ and $\xi_{\text{FL}}$ are fairly independent of $t_{\text{SH}}$ and $t_{\text{FM}}$. This is typically a reasonable assumption for $t_{\text{FM}} > 1$ nm,[18,35] but not the case for $t_{\text{FM}}$ in the ultrathin limit, where $\xi_{\text{FL}}$ has a strong $t_{\text{FM}}$ dependence.[36]

Let us discuss the potential sources of error in the ST-FMR method. In the first and more frequently used ST-FMR type (so-called *lineshape* ST-FMR), *i.e.* fitting to the equations (10,11), the errorbar comes mainly from the uncertainty of $M_s$, $t_{\text{FM}}$ and $t_{\text{SH}}$. $M_s$ and the magnetically dead layer $t_{\text{FM}}^{\text{dead}}$ of the FM film in the multilayer stack can be determined by vibrating sample magnetometry (VSM) or SQUID. The errorbars of $M_s$ and $t_{\text{FM}}^{\text{dead}}$ are transferred directly to that of $\xi_{\text{FMR}}$, therefore sufficient effort has to be made to reduce them.

The second ST-FMR type (so-called *linewidth* ST-FMR), *i.e.* fitting measured $\Delta$ with varying $J_{\text{SH,dc}}$ to equation (12), has all of the error sources described above for the first type. Although this type, which is based on the anti-damping effect induced by the DC bias current on the ST-FMR linewidth, eliminates the hassle of repeated measurements on many samples, it suffers two additional error sources: (i) the determination of $J_{\text{SH,dc}}$ which requires good estimations of the thickness and conductivity of the SH and FM layers, and (ii) the noise in the $V_{\text{mix}}$ trace, when not microwave engineered well enough, causes substantial errorbar in $\Delta$. Furthermore, if the DL ST is not strong enough, $\Delta$ variation vs $J_{\text{SH,dc}}$ can be too small for a good linear fit. This ST-FMR type is therefore most suitable for samples having high anticipated DL ST efficiency.

There is another, less popular, type of ST-FMR[37]: estimate the spin current density directly from the symmetric component of the $V_{\text{mix}}$ lineshape, then divide it with the calculated electrical current density in the SH layer $J_{\text{SH,rf}}$. This is mostly used for sample with highly resistive SH layer (e.g. TIs) in which case most of the RF current is shunted into the FM layer, resulting in a small Oersted field, hence small asymmetric component of the $V_{\text{mix}}$ lineshape and large $S/A$ errorbar. The type has all the aforementioned error sources and an additional one: the transmission coefficient of the RF waves from the source to the sample which has to be carefully calibrated for every set of $I_{\text{rf}}$ and $f$ used.

So far in the analysis, we ignore the contribution of the spin pumping effect from the FM to the SH layer which results in an additional contribution to the symmetrical part of the readout voltage via the ISHE the same way as in the Spin pumping technique described in the previous section. This extra contribution causes substantial overestimation of the DL ST efficiency when the AMR effect of the FM layer (e.g. CoFeB) is sufficiently small, or when the DL ST efficiency is very large (e.g. in W). In such cases, the contribution of the spin pumping effect can be corrected in the analysis[38] or avoided by rotating the applied field in $xz$ plane.[39]

Finally, we note that ST-FMR is mostly used for in-planed magnetized samples, as described above. Modified ST-FMR versions for perpendicularly magnetized samples has been actively pursued.[40,41] Regarding the sample size, Xu *et al.*[42] has shown that ST-FMR gives consistent results for sample widths larger than 400 nm.

## VI. DC current switching

The DC current switching of SOT-MRAM[10] is a preliminary assessment on the performance of the SOT-MRAM prior to more industry standard testing with nanosecond pulses. A number of valuable information can be extracted: switching current, DL ST efficiency and thermal stability factor. Its main drawback is the nanofabrication of SOT-MRAM which is notoriously challenging and costly for most research groups and not yet possible for many exotic SOT materials (e.g. TIs and TMDs).

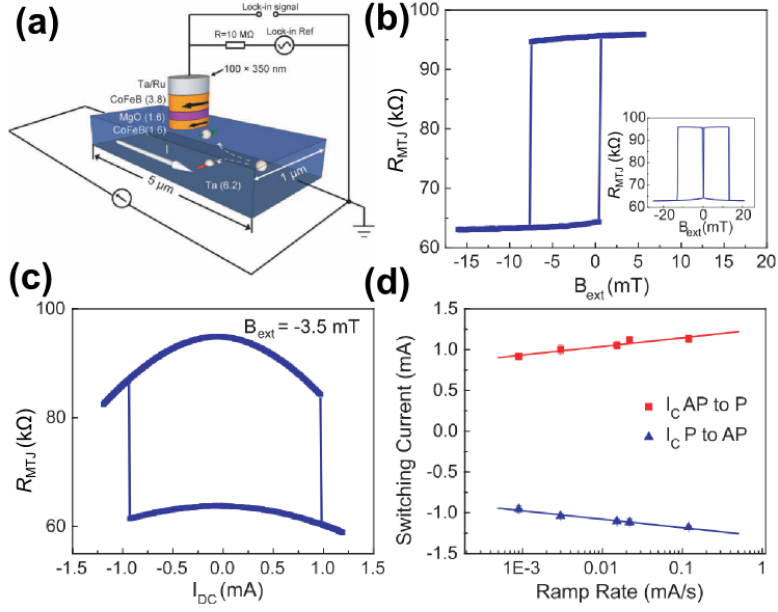

Figure 6: DC current switching of SOT-MRAM. (a) Schematic of a SOT-MRAM device and circuit for measurements. (b) Magnetic minor loop of the MTJ as a function of the external field applied in-plane along the long axis of the sample. (Inset) Magnetic major loop of the device. (c) MTJ resistance as a function of applied DC current. (d) Switching currents as functions of the ramp rate of the sweeping current. Solid lines represent linear fits of switching current versus log(ramp rate). Courtesy of Liu *et al.*[10]

As depicted in Fig. 6(a), the device consists of a magnetic tunnel junction (MTJ) placed on top of a SH channel. The MTJ needs to be a nano-scaled ellipse to achieve uniform magnetization (single domain) and minimize micromagnetic effects.[43,44] Although other configurations are possible,[45] we consider the simplest case[10] in which the MTJ is in-plane magnetized and has its magnetic easy axis (geometrically major axis) placed perpendicularly to the applied current (in the channel). An in-plane constant field may be needed to set the MTJ magnetization near the center of the magnetic minor loop shown in Fig. 6(b). The switching current $I_c$ from AP to P state and vice versa (P: parallel, AP: antiparallel) is measured by sweeping the channel current at a ramp rate $dI/dt$, as shown in Fig. 6(c).

According to the thermally activated switching model,[46,47] $I_c$ depends on the current ramp rate $dI/dt$ as

$$I_c = I_0 \left[ 1 + \frac{1}{\Delta_g} \ln \left( \frac{dI/dt}{I_0} \tau_0 \Delta_g \right) \right], \quad (14)$$

where $\Delta_g = E_g/k_B T$ is the thermal stability factor ($E_g$: barrier energy, $k_B$: Boltzmann constant, and $T$: temperature), $\tau_0$ the thermal fluctuation time (typically 1 ns), and $I_0$ the switching current (corresponding to current density $J_0$) at zero thermal fluctuation. By fitting $I_c$ versus $\ln(dI/dt)$, as shown in Fig. 6(d), $I_0$ and $\Delta_g$ can be estimated. It is noted that if pulsed current can be applied, the pulse-width ($t_{\text{pulse}}$) dependent $I_c$ data can also be fitted by equation (14) to extract $I_0$ and $\Delta_g$, except the term $\ln\left(\frac{dI/dt}{I_0}\tau_0\Delta_g\right)$ is replaced by $\ln\left(\frac{t_{pulse}}{\tau_0}\right)$. The DL ST efficiency is then calculated as[48,49]

$$\xi_{DL} = K_{pre}\alpha t_{FM}(2H_c + H_{eff})/J_0, \quad (15)$$

where $H_c$ is the coercive field determined by half of the width of the magnetic minor loop. Note that the determination of $K_{pre}$, $\alpha$ and $H_{eff}$ of the MTJ free layer requires extra experiments such as VSM/SQUID and FMR.

The errorbar of $\xi_{DL}$ in this method comes from two main sources: the estimation of $K_{pre}$ and $t_{FM}$ (which are not straightforward to measure as discussed in previous sections) and the linear fitting to equation (14). The latter determines $\Delta_g$, which is an important parameter for MRAM applications (related to memory retention). Therefore, it is important to minimize the error of the linear fitting of equation (14) by spanning the ramp rate $dI/dt$ over a few decades and measuring $I_c$ multiple times for each ramp rate.

There are some variations to the technique, mainly to avoid the hassle of nanofabricating the MTJ. One of them is the combination of a Hall bar and a nanopillar acting as the free layer, demonstrated by Mihajlovic et al.,[50] in which the state of the free layer is measured by the Hall voltage via the planar Hall effect (PHE). Another one is using the full Hall bar as the free layer, demonstrated by Liu et al.,[51] which eliminates the need for e-beam lithography but may suffer error due to multi-domain magnetic reversal.[43]

It is also important to note that for magnetic heterostructures with perpendicular magnetic anisotropy (PMA), current-induced SOT switching measurements are much easier to perform for two reasons: (1) Micron-sized Hall bar devices can be adopted to probe magnetization through anomalous Hall effect (AHE), and (2) Even in micron-sized devices, current-induced magnetization switching can still be achieved via domain nucleation and SOT-driven domain wall propagation with an additional in-plane field applied[44]. The simple measurement scheme and much easier fabrication process made this type of switching measurement the most popular approach to demonstrate the efficacy of SOT. However, the multidomain nature of such switching dynamics further complicates the DL ST efficiency estimation from the obtained data. Nevertheless, $\xi_{DL}$ can be more legitimately estimated from current switching data for PMA devices provided that the FM layer being single domain (nano-scaled pillar) and therefore the switching process is coherent as demonstrated by Lee et al.[52]

## VII. Harmonic Hall technique

The AC harmonic Hall technique (sometimes shortened as the harmonic technique) is probably the most popular method used in the community at the time of this writing thanks to its many advantages: easy sample preparation, simple AC measurements at low frequency and estimation of both the DL and FL torques. The sample is typically a micron-sized Hall bar from a SH/FM bilayer, in which the FM has PMA. An external field $H_{ext}$ and an AC current (of a few kHz) are applied to the Hall bar, and the 1st and 2nd harmonic responses of the Hall voltage are read out. The AC signal generation and detection can be most effectively achieved by using a standard lock-in amplifier. By cleverly varying $H_{ext}$ (strength, direction) and analyzing the Hall voltage responses, one can extract the longitudinal and transverse effective fields (induced by the DL and FL torques) which are then converted to ST efficiencies by equation (3). The difficulty of the technique is the segregation of many magnetoresistance as well as thermoelectric effects coming alongside with SOTs.

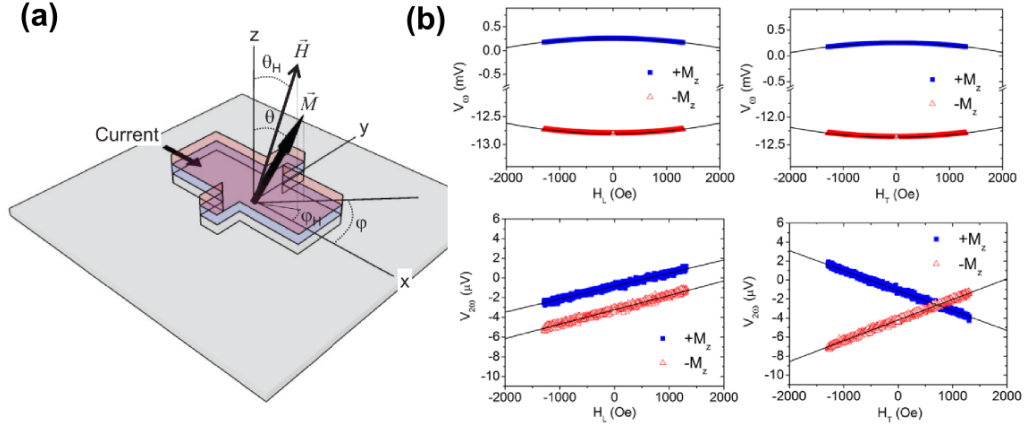

Figure 7: Harmonic Hall technique. (a) Schematic of the sample and experiment. Courtesy of Hayashi et al.[53] (b) Exemplary first and second harmonic Hall traces as functions of sweeping field in longitudinal (L) and transverse (T) direction, for a PMA W/Hf/CoFeB sample. Courtesy of Pai et al.[54]

The harmonic approach to quantify SOT effects was first introduced by Pi et al.[55] Here we describe a simple and more commonly-employed version of the technique, first proposed by Kim et al.[56] and later fully developed by Hayashi et al.[53] This version considers the DL and FL SOTs (represented by the ST effective fields), the anomalous (AHE) and planar (PHE) Hall effect in the sample. The external field is swept along the longitudinal and transverse axes, and the first $V_\omega$ and second $V_{2\omega}$ harmonic responses of the Hall voltage are read out. For a PMA sample whose magnetization direction stays close to the normal of the sample plane, $V_\omega$ is expected to vary with $H_{ext}$ as a cosine function or approximately a quadratic function near $H_{ext} = 0$ with curvature $\zeta = \partial^2 V_\omega / \partial H_{ext}^2$, and $V_{2\omega}$ linearly with $H_{ext}$ with slope $\beta_{L(T)} = \partial V_{2\omega}/\partial H_{ext}^{L(T)}$ (for longitudinal and transverse field), as depicted in Fig. 7. The DL effective field is calculated as

$$H_{DL} = -\frac{2}{\zeta} \cdot \frac{\beta_L \pm 2p\beta_T}{1-4p^2}, \quad (16)$$

where the $\pm$ sign corresponds to the magnetization pointing along $\pm z$-axis. The FL effective field $H_{FL}$ is calculated similarly by interchanging the subscript L and T on the right-hand side of equation (16). The constant $p$ is the ratio of the PHE and AHE resistances which can be measured by rotating the applied field in the $xy$ and $xz$ plane, respectively.

The simplicity and cost-effectiveness of this method make it attractive and popular in the SOT community. The uncertainty of the results comes from usual sources that are required for a good $H_{\text{DL/FL}}$ to $\xi_{\text{DL/FL}}$ conversion via equation (3): the FM's magnetization $M_s$, thickness $t_{\text{FM}}$ and dead layer, and the shunted current density in the SOT layer $j_e$. There exists, however, a puzzling problem with the value of $p$. For most SH/FM bilayers investigated so far, $p$ is very small <0.1. In some cases, notably Pt/Co, it can be as large as 0.3 that the resulting $\xi_{\text{DL}}$ is roughly 2 times higher than when assuming $p = 0$.[57] While there is no consensus on whether and how to take $p$ into account, it is currently advisable that the technique is used when $p$ is sufficiently small that it does not make a big impact.

Note that the analytical expressions in the method's derivation are only valid for a small variation of the magnetization vector and a negligible in-plane anisotropy (compared to the PMA). If, for example, the applied magnetic fields are comparable to or exceeding the anisotropy field that causes a significant deviation of the magnetization vector to the film normal, then the first $V_\omega$ and second $V_{2\omega}$ harmonic responses are no longer quadratic and linear with respect to the applied fields, respectively. Several works have developed more comprehensive models to deal with this issue, such as Garello et al.[58] and Schulz et al.[59] The harmonic Hall voltage method can also be adopted for SH/FMI bilayer devices, where FMI represents ferrimagnetic insulator with PMA (for example, thulium iron garnet, TmIG). However, since the Hall signals therein are mainly originated from the PHE-like and AHE-like spin Hall magnetoresistance (SMR), fitting protocols differ from the SH/FM case are applied to extract $\xi_{\text{DL/FL}}$.[60]

It is also important to note that the harmonic Hall voltage approach is not limited to samples with PMA, albeit less popular. A thorough study on in-plane magnetized samples, which examines the harmonic voltage contributions from DL torque, FL torque, and thermoelectric effects (such as anomalous Nernst effect, ANE), has been given by Avci et al.[61] As shown in Fig. 8(a), an $H_{\text{ext}}$ is applied and swept in the $xy$ plane for 360° to obtain the angle dependence of second harmonic Hall voltage $V_{2\omega}(\varphi)$ (or equivalently the Hall resistance $R_{xy}^{2\omega}(\varphi)$), which can be fitted to

$$V_{2\omega} = \left[V_{\text{AHE}}\left(\frac{H_{\text{DL}}}{H_{\text{ext}}-H_{\text{eff}}}\right) + V_{\text{ANE}}\right]\cos\varphi + \left[2V_{\text{PHE}}\left(\frac{H_{\text{FL}}+H_{\text{Oe}}}{H_{\text{ext}}}\right)\right](2\cos^3\varphi - \cos\varphi) \quad (17)$$

to extract the $\cos\varphi$ dependent term (from $H_{\text{DL}}$ and ANE) and the $2\cos^3\varphi - \cos\varphi$ dependent term (from $H_{\text{FL}}$ and the Oersted field $H_{\text{Oe}}$). Next, $H_{\text{DL}}$ (with $V_{\text{ANE}}$), and $H_{\text{FL}}$ (with $H_{\text{Oe}}$) can be estimated from the linear fits of $H_{\text{ext}}$-dependent results of above-mentioned terms, as shown in Fig. 8(b), from which $\xi_{\text{DL/FL}}$ can be further calculated ($V_{\text{AHE}}$ and $V_{\text{PHE}}$ are the first harmonic AHE and PHE voltages, respectively).

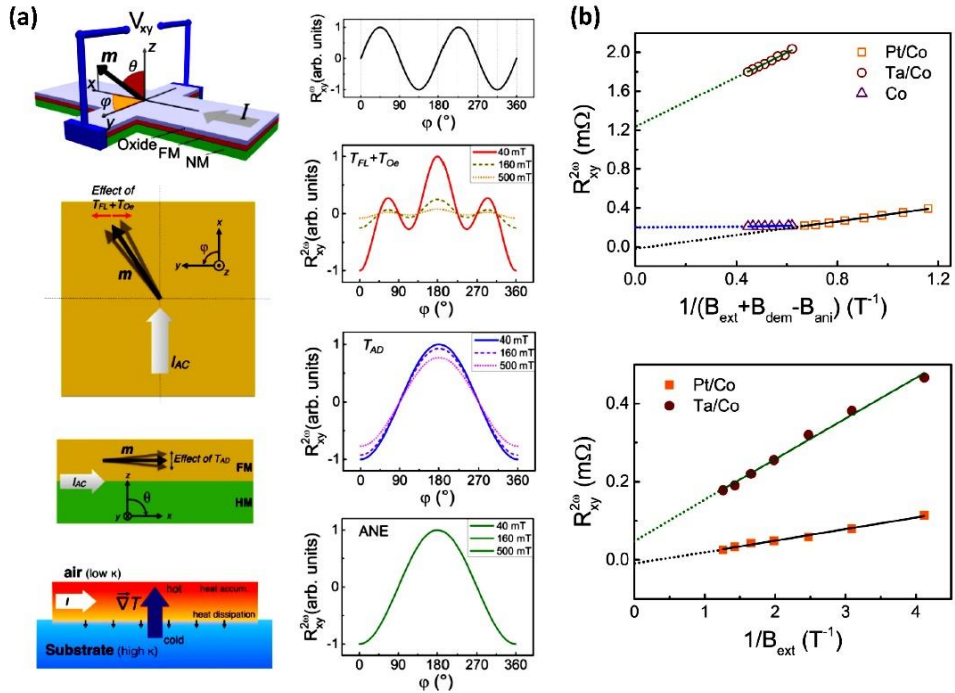

Figure 8: Harmonic Hall technique for in-plane magnetized samples. (a) Schematic of the experiment and harmonic signal contributions, including the effects from FL, Oersted-field, DL torques, and ANE. (b) Field dependence of the second harmonic signal contributions from different origins (DL torque plus ANE vs. Oersted field plus FL torque). Courtesy of Avci et al.[61]

## VIII. Hysteresis loop shift

All the techniques mentioned in previous sections rely on the assumption that the magnetization in ferromagnetic layers can be treated as a macrospin vector. However, in reality, when the prepared samples are in micrometer scales, the magnetization switching and dynamics typically involve magnetic domain nucleation and domain wall motion, rather than a single domain behavior. In fact, two seminal works on the current-induced SOT-driven domain wall motion have shown that not only the SHE but also the interfacial Dzyaloshinskii-Moriya interaction (DMI, which determines the chirality of the domain wall moments) influence the current-driven domain wall dynamics.[62,63] Later, Lee et al.[44] identified that this SHE+DMI scenario should be adopted to explain the necessity of applying an in-plane magnetic field to achieve deterministic SOT switching in samples with PMA.

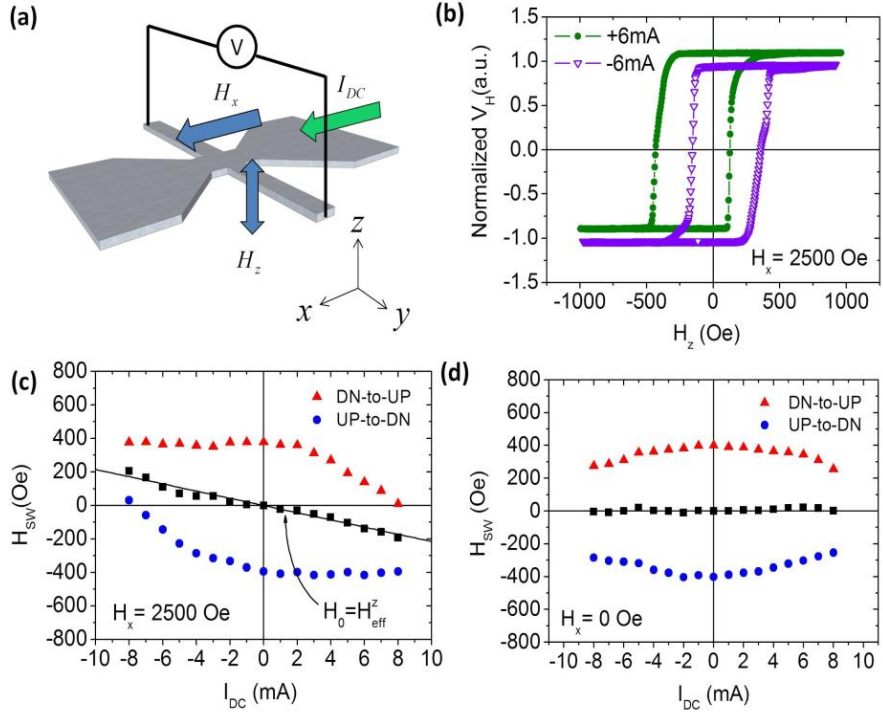

Figure 9: Hysteresis loop shift method. (a) Experiment setup. (b) Representative AHE loop shift results for a PMA Pt/Co/MgO sample with $H_x$ = 2500 Oe and $I_{DC}$ = ±6 mA. (c) $H_{eff}^z$ (as determined by the center of AHE loops) as a function of $I_{DC}$ for $H_x$ = 2500 Oe. (d) Absence of $I_{DC}$-dependent loop shift for $H_x$= 0 Oe. Courtesy of Pai et al.[64]

Combining these concepts, Pai et al.[64] formalized a simple current-induced hysteresis loop shift measurement (loop shift measurement for short) that can simultaneously quantify the DL ST efficiency and the magnitude of the DMI effective field. The sample is typically a micron-sized Hall bar made from a SH/FM multilayer with PMA for AHE signal detection, similar to the Harmonic Hall technique described in the previous section. Two directions of magnetic fields are required, as depicted in Fig. 9. The static in-plane field $H_x$ is used to reorient the domain wall moments that have been affected by the DMI. An out-of-plane hysteresis loop is obtained by AHE readout with sweeping $H_z$ and the application of a DC current $I_{DC}$. Under this circumstance, during magnetization switching, the DL ST or its corresponding $H_{DL}$ acting upon the domain wall moments will give rise to a shift in the detected hysteresis loop, which is denoted as $H_{eff}^z$. Note that this shift should be linearly proportional to the applied current density and $I_{DC}$. When the DMI effective field is fully overcome by $H_x$, $H_{DL}$ can be calculated as $(2/\pi) \cdot H_{eff}^z$. Equation (3) then can be used to estimate the DL ST efficiency $\xi_{DL}$ of the sample. Note that the $2/\pi$ factor is to account for the domain wall motion nature of the ST-driven switching.[65] A direct domain wall motion detection approach (rather than relying on AHE measurement) to quantify DL ST efficiency was proposed by Emori et al.[66]

Due to the simplicity and the DC nature of its electrical measurement, where AC capability such as lock-in amplifier is not required, the technique gained significant popularity after it was formalized. However, the loop shift technique is only applicable for samples with PMA, and the FL ST term ($\xi_{FL}$) cannot be captured. By adopting loop shift method, the uncertainty for $\xi_{DL}$ determination originates from the same sources as in the Harmonic Hall approach. Note that, for both loop shift and harmonic voltage techniques, the Hall bar device lateral geometry can introduce additional uncertainties due to the overestimation of the applied current density flowing in the device, as pointed out by Tsai *et al.*[67] and Neumann *et al.*[68]

## IX. Spin Hall magnetoresistance

After the discovery of the SHE in transition metal systems in early 2010s, several novel magnetoresistance effects based on the theories of SHE and SOT have also been discovered in various SH/FM heterostructures, notably among them is the spin Hall magnetoresistance (SMR). Although there are several variations, the earliest form of SMR was theorized[17] and experimentally observed[69] in SH/FMI bilayer structures, where FMI stands for ferrimagnetic insulator, such as yttrium iron garnet (YIG).

As illustrated in Fig. 10(a), the concept of SMR is related to the spin transmission at the SH/FMI interface: When a longitudinal charge current is applied to a SH/FMI device, all the charge current will be flowing in the SH metal (*e.g.* Pt) since the FMI layer (*e.g.* YIG) is insulating. However, due to the SHE, a transverse spin current will develop, flowing normal to the SH/FMI interface. If the spin polarization direction $\hat{\sigma}$ of this spin current is not collinear with respect to the magnetization direction *m* in the FMI, spin torque transfer would occur and therefore the spin current will be transmitted. In contrast, if $\hat{\sigma}$ is collinear to *m*, then no spin transmission would occur and thereby causing a spin current reflection back into the SH metal layer, resulting in an additional charge current due to the ISHE. Therefore, the resistance of the device will be lower (higher) as $\hat{\sigma}$ is collinear (perpendicular) to *m*. Note that the SMR effect originates from both SHE and ISHE.

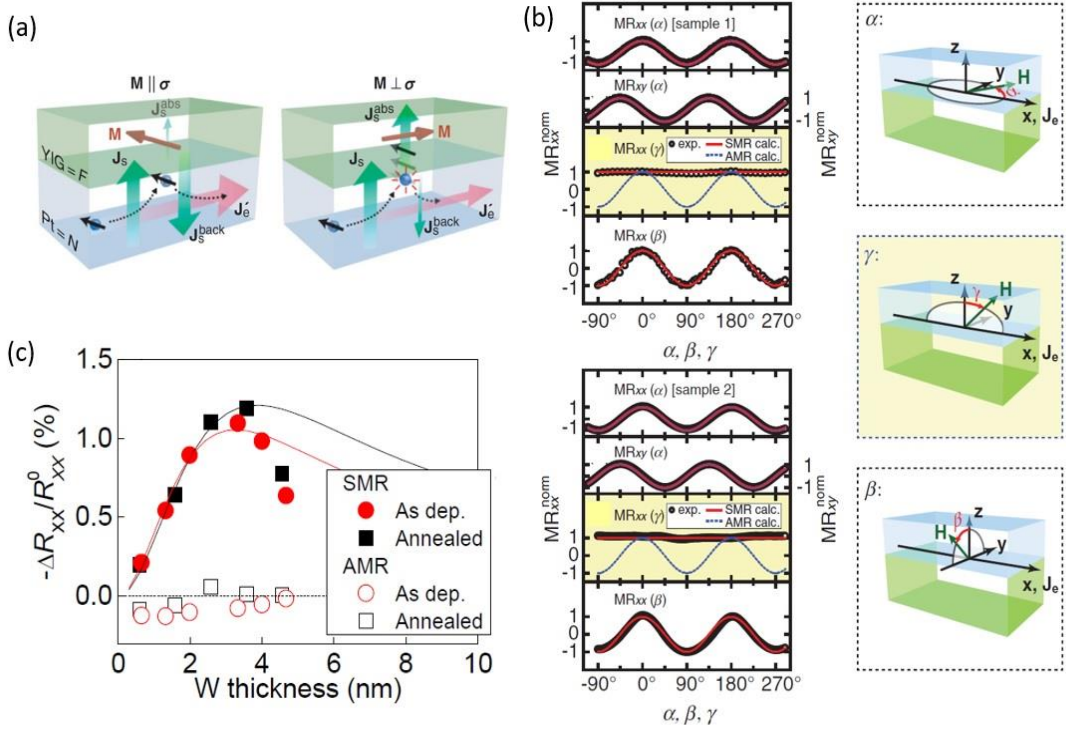

Figure 10: Spin Hall magnetoresistance technique. (a) Illustration of SMR and (b) representative data obtained from Pt/YIG devices. Courtesy of Nakayama et al.[69] (c) SH metal (W) thickness dependence of SMR ratio from a series of W/CoFeB/MgO devices, from which eq. (18) can be fitted to estimate the magnitude of $\theta_{SH}$ and $\lambda_{SH}$. Courtesy of Kim et al.[70]

A typical SMR measurement is done by measuring the longitudinal resistance of a SH/FMI bilayer device while rotating the magnetic field direction in the $xy$-plane or $yz$-plane ($xz$-plane will show no SMR response), as depicted in Fig. 10(b). Chen et al. provided a simple model to predict the resistance variation,[17] in which the relation between the internal spin Hall angle $\theta_{SH}$ and the measured SMR ratio can be approximately expressed as

$$\frac{\Delta \rho_{xx}}{\rho_{xx}} \approx \frac{\theta_{SH}^2}{t_{SH}/\lambda_{SH}} \text{Re}(T_{int}) \tanh(t_{SH}/2\lambda_{SH}), (18)$$

where $T_{int}$ is defined by equation (4). Using equation (18), one can extract $\lambda_{SH}$ and the magnitude of $\theta_{SH}$ from a data set of SMR vs. $t_{SH}$, for example, see Fig. 10(c) (assuming a large $\text{Re}(G_{mix}) \sim 10^{15} \Omega^{-1} m^{-2}$). Note that if one measures the transverse resistivity (Hall resistivity) of the device, then the SMR will be correlated to the PHE and the AHE components through $\text{Re}(T_{int})$ (DL ST contribution) and $\text{Im}(T_{int})$ (FL ST contribution), respectively.

The advantage of SMR approach is its simplicity: conventional DC longitudinal resistance measurement with a rotatable magnetic field (large enough to align $\boldsymbol{m}$) is sufficient. It is not difficult to find that the errorbar of spin Hall angle estimation mainly comes from the uncertainty of $\text{Re}(G_{mix})$ whose value can be calculated theoretically or obtained from a separate FMR measurement. Most importantly, since SMR originates from both SHE and ISHE, one can only extract the magnitude of $\theta_{SH}$ from the $\theta_{SH}^2$ term, but unable to determine the sign of it. It is worth mentioning that other mechanisms might also give rise to a similar MR, for example, magnetic proximity effect[71] and surface spin-orbit scattering.[72] There also exist several similar spin-dependent MR effects in SH/FM all-metallic heterostructures, such as the sizable SMR in samples with PMA[70,73] and unidirectional SMR (USMR) in samples with in-plane magnetic anisotropy.[74] However, the existence of the texture-dependent anisotropic interfacial magnetoresistance (AIMR) also needs to be carefully considered for a reasonable estimation of the SMR, as recently suggested by Philippi-Kobs *et al.*[75]

## X. Discussion

In the previous sections, we revisit seven popular techniques to quantify the SOTs in SH/FM heterostructures with their differences in working mechanisms, sample geometries and anisotropies, frequencies of the applied current and orientations of the applied fields. The selection of suitable techniques for a particular sample depends on the material properties, fabrication capabilities and lab equipment. For example, spin pumping measurement requires a minimal device fabrication capability, but a decent RF signal source to excite ferromagnetic resonance and careful calibrations for precise SOT efficiency quantification. In contrast, costly and time-consuming nano-fabrication processes are unavoidable for direct SOT switching measurement yet the obtained data can be directly applicable for SOT-based device benchmarking. In terms of materials, harmonic voltage and loop shift approaches can be readily deployed for SH/FM heterostructures with PMA, whereas spin pumping and ST-FMR are more widely-used for in-plane magnetized samples, and SMR can be used for SH/FMI bilayers. A summary and comparison of the techniques covered in previous sections are provided in Table 1.

Table 1: Summary of the popular SOT quantification techniques covered in this article.

| Technique | Concept | Measure | Device | Cost | Critical parameters | Pros | Cons |
| --- | --- | --- | --- | --- | --- | --- | --- |
| NLSV | SHE, ISHE, lateral spin diffusion | DL ST efficiency | Nano-sized NLSV | High | signal-to-noise ratio, device geometry | Direct observation of SHE and ISHE | Fab-intensive, often requires low temperatures |
| Spin pumping | FMR-driven ISHE | DL ST efficiency | Micron-sized or mm-sized bar | Medium | Precession cone angle, resistivities | Fab-light, simple, and versatile | Correction factor calibrations required, geometry or setup-induced artefacts |
| ST-FMR | SOT driven FMR (AMR) | DL & FL ST efficiency | Micron-sized bar | Medium | $M_s$, $t_{dead}$ | Fab-light, self-calibrated, and versatile | Hard to disentangle FL torque |

| Current switching | SOT-driven magnetic reversal | DL ST efficiency | Nano-sized MTJ | High | $M_s$, $t_{dead}$ | Close to applications | Fab-intensive, influence of micromagnetics |
|---|---|---|---|---|---|---|---|
| Harmonic Hall | SOT-modulated Hall voltages | DL & FL ST effective fields | Micron-sized Hall bar | Low | $M_s$, $t_{dead}$, Hall bar geometry | Fab-light, simple, and quick | Hard to disentangle other harmonic signal contributions |
| Loop shift | SOT-assisted magnetic reversal | DL ST & (magnitude of) DMI effective fields | Micron-sized Hall bar | Low | $M_s$, $t_{dead}$, Hall bar geometry | Fab-light, DC measurement | Cannot capture FL torque |
| SMR | SHE, ISHE, and SOT | (magnitude of) spin Hall angle | Micron-sized Hall bar | Low | Spin-mixing conductance | Fab-light, DC measurement | Cannot determine the sign of spin Hall angle, hard to disentangle other MR signals |

As discussed in section II, all of the aforementioned SOT quantification methods make several assumptions in the magnetization dynamics, SOT effects, and spin transport in the sample. Notably, the macrospin model works well in essentially uniform samples but may break down in some corner cases. One of them is the magnetic reversal of in-plane magnetized nanopillars in which the magnetic profile near the edges plays an important role. As shown by micromagnetic simulations[43], the nanoscale SOT-MRAM magnetic reversal depends strongly on the edge roughness, domain nucleation as well as the Oersted field induced by the current. The role of domain depinning and multidomain dynamics in magnetic reversal is more profound in PMA samples.[44,76] The error caused by these domain-related effects to the macrospin model in magnetic reversal is not well calibrated, but it is advisable to keep it in consideration of the results.

Note that in the analyses described above, we assume that the SH layer's thickness $t_{\text{SH}}$ is a few times larger than its spin diffusion length $\lambda_{\text{SH}}$ so that the transverse spin current density at the SH/FM interface $j_s^{\text{SH/FM}}$ is related to the electrical current density $j_e$ in the SH layer by equation (1). When the SH thickness is within a couple of spin diffusion length, however, because of spin reflection at the other surface of the SH layer, the relationship becomes (modified from [77]):

$$j_s^{\text{SH/FM}} = \left(\frac{\hbar}{2e}\right)\theta_{\text{SH}} j_e (1 - \text{sech}(t_{\text{SH}}/\lambda_{\text{SH}})). \quad (19)$$

Experimentally, this appears as thickness dependent $\xi_{\text{SH}}$ which was observed by many experiments. Note that similar argument applies when $t_{\text{FM}}$ is sufficiently small as well (as in MTJ), although the analysis is more complicated. Furthermore, recall that $\xi_{\text{SH}} = \theta_{\text{SH}} T_{\text{int}}$ where the interfacial spin transmission factor $T_{\text{int}}$ also depends on $t_{\text{SH}}$ as described in equation (4) (which already includes eq. (19) above). Thus, great care must be given when $t_{\text{SH}}$ is small (which is often desirable in applications).

In some special cases, $j_e$ uniformity needs to be considered. Because the resistivity of the SH layer is higher near the SH/FM interface due to surface scattering,[24] the current distribution near the interface is nonuniform. The uniformity of SHE-induced spin current depends on the SHE mechanism[13] of the SH material. If the SHE mechanism is skew-scattering, $\theta_{SH}$ does not depend on $\rho_{SH}$, hence is constant throughout the SH layer. Because $j_e$ is nonuniform, so is $j_s$. If the SHE mechanism is intrinsic or side-jump, $\theta_{SH} \propto \rho_{SH}$, but $j_e \propto 1/\rho_{SH}$ (for constant electric field), so $j_s$ is uniform. In either case, $\lambda_{SH}$ can be nonuniform which requires elaborate modification of equation (19).[25] To avoid such hassles, it is advisable to have a thick enough SH layer for consistent results. Furthermore, for systems with interfacial origin of the SOT, the $\xi_{SH}$ might have a different thickness dependence from equation (19). Replacing FM layers with antiferromagnetic (AFM) materials will also complicate the analysis, for example see Chiang *et al.*[78] Lastly, lateral geometry of the sample should also be carefully considered to avoid nonuniform current distribution that may lead to poor SOTs estimation as pointed out by Neumann *et al.*[68]

It is obvious that the list of seven techniques reviewed in this article is by no means exhaustive. There have been many other techniques developed over the last decade, although are less popular but utilize interesting phenomena in probing SOTs or estimating the charge-spin conversion efficiency. Notable among them are the SHE tunneling spectroscopy proposed by Liu *et al.*,[79] resonance spectroscopy by Emori *et al.*,[80] longitudinal magneto-optic Kerr effect (MOKE) method by Fan *et al.*,[36] SHE-driven chiral domain wall motion technique by Emori *et al.*,[66] and thermal injection method by Qu *et al.*[81]

As a final remark, we address the challenges that these existing quantification techniques might face while the systems-of-interests are heterostructures with emergent materials such as TIs, TMDs, and Weyl semimetals. First of all, unlike conventional metallic SH layers, these exotic materials, such as $Bi_2Se_3$,[37,82–84] $(Bi_{0.5}Sb_{0.5})_2Te_3$,[85,86] $WTe_2$,[87,88] and $MoTe_2$[89–91] are typically highly resistive ($\rho_{SH} > 10^2 \mu\Omega$-cm). The huge current distribution imbalance between the SH and the FM layers might undermine the accuracy of SOT quantification. For example, one can arrive at a humongous $\xi_{SH}$ even if the measured $H_{DL}/I_e$ is small, provided a large enough SH layer resistivity. The small error in $H_{DL}/I_e$ quantification could propagate and become a large one in $\xi_{SH}$ estimation. The possibility of having surface state contribution[92,93] or unconventional torques[87,94,95] would further complicate the quantification. Parasitic thermoelectric effects can also be significant in these materials and therefore one should be more cautious about possible artefacts. For example, it has been shown that the harmonic voltage signals might be plagued by the Nernst effect in TIs or semimetals,[96] which could potentially lead to overestimations of $\xi_{SH}$ for such SH layers. Disentangling various spurious contributions from the real SOT effects in versatile quantification techniques will be essential for meaningful estimation of the SOTs and charge-spin conversion efficiencies in these new categories of materials.

**Abbreviations**

| | |
|---|---|
| AC | Alternating current |
| AFM | Antiferromagnetic |
| AHE | Anomalous Hall effect |
| AMR | Anisotropic magnetoresistance |
| CPW | Coplanar waveguide |
| DC | Direct current |
| DL | Damping-like |
| DMI | Dzyaloshinskii-Moriya interaction |
| FL | Field-like |
| FM | Ferromagnet / Ferromagnetic |
| FMI | Ferrimagnetic insulator |
| FMR | Ferromagnetic resonance |
| ISHE | Inverse spin Hall effect |
| MOKE | Magneto-optic Kerr effect |
| MRAM | Magnetoresistive random access memory |
| MTJ | Magnetic tunnel junction |
| NLSV | Nonlocal spinvalve |
| P/AP (MTJ state) | Parallel / Antiparallel |
| PHE | Planar Hall effect |
| PMA | Perpendicular magnetic anisotropy |
| RF | Radio frequency |
| SH | Spin Hall |
| SHE | Spin Hall effect |
| SMR | Spin Hall magnetoresistance |
| SOT | Spin-orbit torque |
| SQUID | Superconducting quantum interference device |
| ST | Spin torque |
| STT | Spin transfer torque |
| TI | Topological insulator |
| TMD | Transition metal dichalcogenide |
| USMR | Unidirectional spin Hall magnetoresistance |
| VSM | Vibrating sample magnetometry |

**Symbols**

| | |
|---|---|
| $G_{\text{mix}}$ | Spin-mixing conductance |
| $G_{\text{mix}}^{\text{eff}}$ | Effective spin-mixing conductance |
| $H_{\text{eff}}$ | Effective demagnetization field |
| $H_{\text{ext}}$ | Externally applied magnetic field |
| $H_c$ | Coercive field |
| $K_{\text{pre}}$ | $e\mu_0 M_s/\hbar$ |
| $M_s$ | Saturation magnetization |
| $T_{\text{int}}$ | Interfacial spin transmission coefficient |
| $\boldsymbol{H}_{\text{DL(FL)}}$ | DL(FL) effective field |
| $\boldsymbol{j}_{e(s)}$ | Charge (spin) current density |
| $\theta_{SH}$ | Spin Hall angle |
| $\mu_0$ | Vacuum permeability |
| $\hat{\sigma}$ | Spin-polarization unit vector |
| $\tau_0$ | Thermal fluctuation time |

| | | |
|---|---|---|
| | $\hbar$ | Reduced Planck constant |
| | $\Delta$ | Linewidth |
| | $\Delta_g$ | Thermal stability factor |
| | $\Phi$ | Precession cone angle |
| | $e$ | Electron charge |
| | $t$ | Thickness |
| | $w$ | Width |
| | **$M$** | Magnetization vector |
| | **$m$** | Normalized magnetization vector |
| | $\alpha$ | Gilbert damping coefficient |
| | $\gamma$ | Gyromagnetic ratio |
| | $\lambda$ | Spin diffusion length |
| | $\xi$ | Spin torque efficiency |
| | $\rho$ | Resistivity |
| | $\sigma$ | Electrical conductivity |
| | $\omega = 2\pi f$ | Frequency |

**Acknowledgements**
The authors acknowledge support from the Ministry of Science and Technology of Taiwan (MOST) under grant No. MOST-109-2636-M-002-006.

**Data Availability Statement**
The data that supports the findings of this study are available within the article.